\documentclass{pr-imfp05}

\newcommand{\lsim}{\mathrel{\mathop{\kern 0pt \rlap
  {\raise.2ex\hbox{$<$}}}
  \lower.9ex\hbox{\kern-.190em $\sim$}}}
\newcommand{\gsim}{\mathrel{\mathop{\kern 0pt \rlap
  {\raise.2ex\hbox{$>$}}}
  \lower.9ex\hbox{\kern-.190em $\sim$}}}

\begin{document}

\title{Direct searches for Dark Matter Particles: WIMPs and axions}

\author{Igor G. Irastorza}

\address{CEA, Centre d'Etudes Nucl\'eaires de Saclay, \\DSM/DAPNIA, 91191 Gif-sur-Yvette Cedex, France\\
E-mail: Igor.Irastorza@cern.ch}


\maketitle

\abstracts{WIMPs and axions are the two best motivated candidates
to compose the Dark Matter of the Universe. An important number of
experimental groups are developing and using different techniques
for their direct detection. An updated review of current searches
is done, emphasizing latest results.}

\section{Introduction}

Since first suggested by Zwicky in the 1930s, the existence of an
invisible and unconventional matter as a dominant part of our
Universe has been supported by an ever increasing body of
observational data. The latest precision cosmology measurements
\cite{Spergel:2003cb} further constrain the geometry of the
Universe to be flat ($\Omega \sim 1\pm0.04$), and its composition
(to the level of a few \%) to be mostly dark energy
($\Omega_\Lambda \sim 73\%$) and non-baryonic dark matter
($\Omega_{NB} \sim 23\%$), leaving less than $\sim 4\%$ for
ordinary baryonic matter. Dark energy is a theoretical concept
related to Einstein's cosmological constant, the nature of which
is essentially unknown. Dark matter, on the contrary, could be
composed by elementary particles with relatively known properties,
and which could be searched for by a variety of means. These
particles must have mass, be electrically neutral and interact
very weakly with the rest of matter. They must provide a way of
being copiously produced in the early stages of the Universe life,
so they fill the above-mentioned $\sim$23\% of the Universe
contents. Neutrinos are the only standard particles fitting in
that scheme, but the hypothesis of neutrinos being the sole
component of dark matter fails to reproduce part of the
cosmological observations, in particular the current structure of
the universe. The dark matter problem is therefore solved only by
going into models beyond the standard model of elementary
particles, among which two generic categories emerge as the best
motivated for the task: WIMPs and axions.

WIMP is a generic denomination for any Weakly Interacting Massive
Particle. A typical example of WIMP is the lightest supersymmetric
particle (LSP) of SUSY extensions of the standard model, usually
the neutralino. They would have been thermally produced after the
Big Bang, cooled down and then frozen out of equilibrium providing
a relic density\cite{Primack:1988zm,Jungman:1995df}. The
interesting mass window for the WIMPs spans from a few GeV up to
the $\sim$ TeV scale, but can be further constrained for specific
models and considerations.

Axions, on the contrary, are light pseudoscalar particles that are
introduced in extensions of the Standard Model including the
Peccei-Quinn symmetry as a solution to the strong CP
problem\cite{Peccei:1977hh}. This symmetry is spontaneously broken
at some unknown scale $f_a$, and the axion is the associated
pseudo-Goldstone boson \cite{Weinberg:1977ma,Wilczek:1977pj}. The
axion framework provides several ways for them to be produced
copiously in the early stages of the Universe, which makes it a
leading candidate to also solve the dark matter
problem\cite{Raffelt:1990yz,Turner:1989vc}.

The hypothesis of axions or WIMPS composing partially or totally
the missing matter of the Universe is specially appealing because
it comes as an additional bonus to what these particles were
originally thought for, i.e. they are not designed to solve the
dark matter problem, but they may solve it. In addition, the
existence of WIMPs or axions could be at reach of the sensitivity
of current or near future experiments, and this has triggered a
very important experimental activity in the last years. In the
following pages a review is given of the current experimental
efforts to detect these particles by direct means, i.e. aiming at
their direct interaction with terrestrial detectors. Indirect
methods, like those looking for their decay products in
astronomical or cosmic rays observations may also put constraints
on the properties of these
particles\cite{Edsjo:2005ih,Bergstrom:2000pn}, although they
suffer from extra degrees of uncertainties, like the phenomenology
driving the accumulation of dark matter particles in astrophysical
bodies and their decay into other particles, and they are left out
of the scope of the present review.

\section{WIMP searches}

If WIMPs compose the missing matter of the universe, and are
present at galactic scales to explain the observed rotation curves
of the galaxies, the space at Earth location is supposed to be
permeated by a flux of these particles characterized by a density
and velocity distribution that depend on the details of the
galactic halo model\cite{Belli:2002yt,Copi:2002hm,Green:2003yh}. A
common estimate \cite{Lewin:1995rx} (although probably not the
best one) gives a local WIMP density of 0.3 GeV/cm$^3$ and a
maxwellian velocity distribution of width $v_{rms}\simeq270$ km/s,
truncated by the galactic escape velocity $v_{esc}\simeq650$ km/s
and shifted by the relative motion of the solar system through the
galactic halo $v_0=230$ km/s.

The direct detection of WIMPs relies on measuring the nuclear
recoil produced by their elastic scattering off target nuclei in
terrestrial --usually underground-- detectors\cite{Smith:1988kw}.
Due to the weakness of the interaction, the expected signal rates
are very low ($1-10^{-5} $ c/kg/day). In addition, the kinematics
of the reaction tells us that the energy transferred to the
recoiling nuclei is also small (keV range), which in ionization
and scintillation detectors may be further quenched by the fact
that only a fraction of the recoil energy goes to ionization.
These generic properties determine the experimental strategies
needed. In general, what makes these searches uniquely challenging
is the combination of the following requirements: thresholds as
low as possible, and at least in the keV range; ultra low
backgrounds, which implies the application of techniques of
radiopurity, shielding and event discrimination; target masses as
large as possible; and a high control on the stability of
operation over long times, as usually large exposures are needed.

Even if these strategies are thoroughly pursued, one extra
important consideration is to be noted. The small WIMP signal
falls in the low-energy region of the spectrum, where the
radioactive and environmental backgrounds accumulate at much
faster rate and with similar spectral shape. That makes WIMP
signal and background practically indistinguishable by looking at
their spectral features. If a clear positive detection is aimed
for, then more sophisticated discrimination techniques and
specially more WIMP-specific signatures are needed. Several
positive WIMP signatures have been proposed, although all of them
pose additional experimental challenges. The first one is the
\emph{annual modulation}\cite{Drukier:1986tm} of the WIMP signal,
reflecting the periodical change of relative WIMP velocity due to
the motion of the Earth around the Sun. The variation is only of a
few \% over the total WIMP signal, so even larger target masses
are needed\cite{Cebrian:1999qk} to be sensitive to it. This signal
may identify a WIMP in the data, provided a very good control of
systematic effects is available, as it is not difficult to imagine
annual cycles in sources of background. A second WIMP signature is
the \emph{A-dependence signature}\cite{Smith:1988kw}, based on the
fact that WIMPs interact differently (in rate as well as in
spectral shape) with different target nuclei. This signature
should be within reach of set-ups composed by sets of detectors of
different target materials, although the technique must face the
very important question of how to assure the background conditions
of all detectors are the same. Finally, the \emph{directionality
signature}\cite{Spergel:1987kx} is based on the possibility of
measuring the nuclear recoil direction, which in galactic
coordinates would be unmistakably distinguished from any
terrestrial background. This option supposes an important
experimental challenge and it is reserved to gaseous detectors,
where the track left by a nuclear recoil, although small, may be
measurable.

\begin{figure}[t]
\centering\mbox {\epsfig{file=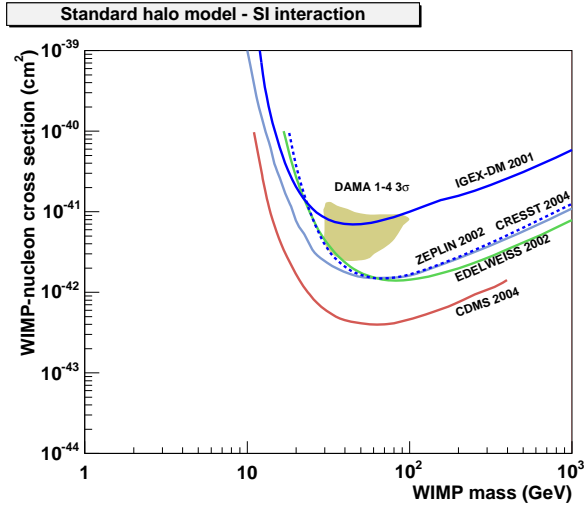,width=8.cm}}
\caption{Exclusion plots of the IGEX-DM, ZEPLIN-I, CRESST,
EDELWEISS and CDMS experiments, as well as the positive region of
the DAMA experiment (only first 4 annual cycles). Standard
assumptions for the halo model and a pure spin-independent
WIMP-nucleon interaction have been considered. See text for more
details.} \label{explot}
\end{figure}

Most of the past and current experiments having given the most
competitive results are not sensitive to any of these positive
WIMP signatures, and their reported results are usually exclusion
plots in the ($\sigma_N,M$) plane, obtained by comparing the total
spectra measured directly with the nuclear recoil spectrum
expected for a WIMP (where $\sigma_N$ is WIMP-nucleon cross
section and $M$ the the WIMP mass). For the sake of comparison
between experiments, these exclusion plots, like the ones in fig.
\ref{explot}, are usually calculated assuming the standard
properties for the halo model previously mentioned, and a spin
independent WIMP-nucleus interaction. For a discussion on how
other halo models or the inclusion of spin-dependent interactions
affects these exclusion plots see
\cite{Belli:2002yt,Copi:2002hm,Green:2003yh}. As will be discussed
in the following pages, the DAMA experiment claims the detection
of an annual modulation in their data, although the interpretation
as a WIMP signal is controversial. Recent progress in the
experimental techniques promises that experiments in the near
future will have wider access to the three positive signatures
previously mentioned in case a WIMP is detected.

In the following a review of the current status of the
experimental WIMP searches is done, focussing on the latest and
most important results and developments. For an exhaustive and
historical listing of experiments we refer to
\cite{Morales:2001qq,Morales:2002ud,Morales:2005ky}.

\subsection{Ionization detectors}

Ionization germanium detectors represent the conventional approach
of a well-known technology, where radiopurity, background
reduction and shielding techniques have been optimized
extraordinarily during the last two decades, in the context of
double beta decay searches. The result released by the IGEX
collaboration in 2001\cite{Morales:2001hj}, shown in fig.
\ref{explot} exemplifies the state-of-the-art in raw background
reduction techniques. The result was obtained with a setup in the
Canfranc Underground Laboratory composed by an ultrapure germanium
detector of 2.1 kg, surrounded by a shielding of ultrapure
components, including an innermost core of 2.5 tons of
2000-year-old archaeological lead forming a 60 cm side cube,
flushed with clean N$_2$ to remove radon and followed by an extra
20 cm of radiopure lead, a cadmium sheet, muon vetos and 40 cm of
neutron moderator. The achieved threshold was 4 keV, and the
background level was 0.21 c/keV/kg/d between 4-10 keV (0.10 in
10-20 keV, 0.04 in 20-40 keV), the lowest raw background level
achieved up-to-date in this energy range.

Improved results using pure ionization detectors requires the
important challenge of further refining the radiopurity and
shielding techniques. The GEDEON\cite{Morales:2005ky} project
proposes the conservative approach of following the same path of
the successful IGEX technology, extended in mass, and further
optimized from the point of view of radiocleanness. More exotic
ideas exist, like the one proposed within the
GENIUS\cite{Baudis:1998gi} project, of submerging the germanium
crystals in liquid nitrogen, which may act as a very pure
shielding. GENIUS-TF\cite{Klapdor-Kleingrothaus:2004bt} is
currently testing the concept. The GERDA\cite{Schonert:2005zn}
project, a high mass germanium experiment, proposed for double
beta decay, has also joined the quest for WIMP searches.

\subsection{Scintillation detectors}

Scintillation detectors have been extensively used for WIMP
detection, specially because they are the technique which provided
the easiest way to large target masses. It was in fact a first
setup of NaI scintillators which first looked for the annual
modulation signature\cite{Sarsa:1996pa}. With no so good prospects
concerning low background capabilities as germanium detectors,
scintillation may however provide a way --although limited-- to
discriminate between nuclear recoils and electron recoils, due to
their slightly different scintillation decay times.

Currently, the DAMA group gathers the expectation of the field
with its claim of observation of an annual modulation signal of
unexplained origin and perfectly compatible with a WIMP of $\sim
52$ GeV and $\sim7.2 \times 10^{-6}$ pb (and standard assumptions
for the halo model). The DAMA experiment in the Gran Sasso
Laboratory, now completed \cite{Bernabei:2003za}, operated 9
radiopure NaI crystals of 9.7 kg each, viewed by two PMT in
coincidence gathering 107731 kg day of statistics and obtaining
evidence for the modulation along 7 annual cycles. The DAMA
positive signal has been ruled out by other experiments in the
standard scenario represented by the exclusion plots of fig.
\ref{explot}. However, in view of the important uncertainties in
the underlying theoretical frameworks and in the galactic halo
models, it is unclear, and a matter of hot discussion, whether all
results are compatible once all uncertainties are taken into
account. It seems that one can always concentrate on a specific
theoretical framework that allows to accommodate both DAMA
positive result and the other exclusion
plots\cite{Bernabei:2003za}. An additional result using the same
target seems to be needed to solve the controversy. The same DAMA
team is already running an enlarged set-up of $\sim250$ kg of NaI
(LIBRA\cite{Bernabei:2005ki}) and will soon deliver the first set
of data. An independent result will come from the ANAIS
experiment\cite{Cebrian:2005kz}, currently in the way of
instrumenting its $\sim100$ kg of NaI in the Canfranc Underground
Laboratory.

Other scintillating materials have been used in the past. Worth to
be noted is the recent result obtained by the KIMS group, using
CsI crystals\cite{kims}. This material offers a higher potential
of discrimination between nuclear and electron recoils when
compared with NaI, due to the enhanced difference between the
scintillation pulse time pattern.

Experiments using liquid noble gases, especially Xenon, should be
classified half way between ionization and scintillation. The
scintillation mechanism in noble gases is very different than in
the previous cases, and allows for an improved discrimination
capability by exploiting the different time patterns of the
scintillation pulses of nuclear and electron recoils or, more
efficiently, by using the ratio charge/light when operating in
hybrid mode. This second possibility is available in two-phase
prototypes, where an electric field is applied to prevent
recombination and to drift the electrons to the gaseous phase
where they are detected (via the secondary luminescence).

Several groups are developing and using noble liquid detectors for
WIMP searches. They have proven that this technique provides good
prospects of radiopurity and background discrimination and
relatively easy scaling-up.
DAMA/Xe\cite{Bernabei:1998ad,Bernabei:2000qn} is among the
pioneers of the technique, originally motivated for double beta
decay searches. The Xenon program carried out by the ZEPLIN
collaboration in the Boulby Mine Laboratory has produced several
prototypes. The ZEPLIN-I\cite{Alner:2005pa} prototype, using 6 kg
of Xenon in pure scintillation mode, has provided a very
competitive result shown in fig. \ref{explot}. Current effort
focuses on the second phase of the experiment, ZEPLIN-II, that
will operate in the two-phase mode. In additon to ZEPLIN, the
groups XENON\cite{Aprile:2005ww} and X-MASS\cite{xmass} work
towards the design and construction of 100 kg prototypes with
position sensitivity. The key question beneath this is the
self-shielding concept, consisting in performing fiducial cuts of
the detector to achieve the maximum signal-to-background ratio,
exploiting the fact that external background will interact
primarily in the outer parts of the detector volume. For future
generation, larger scale detectors this concept may become very
important, as discussed later. Let us mention finally that large
TPCs of liquid Argon have been also proposed, like for example the
WARP\cite{Brunetti:2004rk} collaboration which profits from the
experience of the ICARUS experiment in those techniques.

\subsection{Cryogenic detectors}

Nuclear recoils can also be detected through the heat (phonons)
created in the detector by the recoiling nucleus. This signal is
detectable in calorimeters operating at cryogenic temperatures, to
which a suitable thermometer is attached. At those temperatures,
the released heat produces a temperature raise that can be
measurable.

The main advantage of this technique is that most of the energy of
the interaction is visible and therefore no quenching factor must
be applied. Besides, the phonon signal potentially provides the
best energy resolution and thresholds. On the other hand, however,
the operation of cryogenic detectors is a relatively complex
technique facing many challenges when going for larger exposure
times and masses. For the same reason, radiopurity techniques are
also more difficult to apply.

A reference point in pure cryogenics detectors is the pioneering
work of the Milano group, now leading the CUORE/CUORICINO
experiment\cite{Arnaboldi:2002du,Arnaboldi:2004qy} in the Gran
Sasso Laboratory. The CUORE project, designed to search for the
neutrinoless double beta decay of $^{130}$Te, intends the
construction of an array of 988 TeO$_2$ cryogenic crystals,
summing up $\sim 750$ kg of bolometric mass. A first step of the
project, CUORICINO, is already in operation and involves 62
($\sim$40.7 kg) crystals, by far the largest cryogenic mass in
operation underground. Although background levels are still too
high to provide competitive limits in WIMP detection, important
progress is being made and CUORE may have very good sensitivity to
WIMP annual modulation\cite{Arnaboldi:2003tu}.

However, cryogenic detectors have taken the lead on WIMP searches
because of the possibility of operating in hybrid mode. Due to the
large choice of target materials available to the cryogenic
techniques, and when the material in question is a semiconductor
or a scintillator, the detector could in principle be operated in
hybrid mode, measuring simultaneously the heat and charge or the
heat and light respectively. This strategy has proven to be the
most competitive and efficient in discriminating nuclear recoils
from electron recoils. In fact, cryogenic ionization experiments,
like CDMS\cite{Akerib:2004fq} in the Soudan Underground Laboratory
and EDELWEISS\cite{Sanglard:2005we} in the Modane Underground
Laboratory, have provided the best WIMPs exclusion plots
up-to-date\footnote{Very recently the CDMS collaboration has
presented new preliminary results\cite{seoul} that further improve
the exclusion plot shown in fig.\ref{explot}}, shown in figure
\ref{explot}. Both experiments presently operate prototypes at
$\sim$1 kg scale, and although their raw backgrounds are
relatively high with respect with pure ionization experiments,
they reject more than $\sim$99.9\% of the electron recoils (by
comparing heat and charge signals), reducing the background to
only a few counts, compatible with the expected neutron background
or the misidentification event rate). Both groups currently work
toward increasing the mass of their set-ups and therefore the
available exposure.

Although currently less competitive than heat and charge, the
simultaneous measurement of heat and light is recently presenting
very interesting prospects. The ROSEBUD group first applied it
underground\cite{Cebrian:2003jr} and recently the CRESST
collaboration has presented a very competitive exclusion plot
obtained with two 300 g CaWO$_4$ prototypes\cite{Angloher:2004tr}
(dashed line in fig. \ref{explot}). A very relevant feature of
this result is that tungsten recoils can be distinguished -with
some efficiency- from O or Ca recoils by virtue of their different
ratio heat/light. This improves substantially the sensitivity of
the experiment, as neutrons are expected to interact more with
lighter nuclei, unlike WIMPs. In addition, recent scintillation
studies \cite{Coron:2004iy} have shown that a large variety of
scintillating crystals are available. This opens the way to use
sets of different crystals operating in this mode to look for the
$A$-dependence WIMP signature. While the use of this signal in
conventional detectors suffers from large uncontrolled systematics
derived from the fact that one cannot assure the background to be
the same for different crystals, light/heat hybrid detectors,
sensitive only to neutrons, may overcome this difficulty. In this
line, the ROSEBUD collaboration has successfully operated
underground a set of 3 different bolometers in the same setup
\cite{Cebrian:2004ws}, sharing similar external background
conditions.

\subsection{New approaches and strategies for the future}

Some techniques that fall out of the above classification have
been proposed and some of them are being developed with some
degree of success. They intent to solve some of the problems of
the conventional previous techniques in new, original ways, to
find identificative WIMP signatures or to explore possibilities
less favoured by the most standard theoretical scenarios (like,
for example, cases where the WIMP interacts primarily via
spin-dependent cross sections). Without entering into details,
worth to mention are the superheated droplets detectors, like
SIMPLE\cite{Girard:2005pt}, PICASSO\cite{Barnabe-Heider:2005pg},
or detectors based on superfluid He$^3$, like
MACHe3\cite{Winkelmann:2005du}. New ideas appear continuously, for
example the spherical TPC
concept\cite{Giomataris:2003pd,Aune:2005is}, proposed in the
context of neutrino physics, and which applicability to WIMP
detection is under study.

An important category are the techniques aiming at the detection
of the nuclear recoil direction. As mentioned before, such signal
would suppose a definitive positive signature of a WIMP and would
in addition give information about how they are distributed in the
halo. While being an important technical challenge this
measurement might be performed in gaseous detectors, where nuclei
of $\sim 10-100$ keV could leave tracks in the mm--cm range
(depending on the pressure and nature of the gas). The
DRIFT\cite{Ayad:2003ph} experiment is proving the technique of low
pressure negative ion TPC\cite{Martoff:2000wi}. Low pressure (40
Torr) makes the tracks to be relatively long (few cm), and the
addition of electronegative gas (CS$_2$) makes the electrons to be
captured, so the negative ions drift to the avalanche region (a
multiwire proportional chamber) with much smaller diffusion and no
magnetic field is needed. A first prototype DRIFT-I has already
worked successfully, and its extension to DRIFT-II and DRIFT-III
are envisaged. Recently another group, NEWAGE\cite{newage}, has
joined the development of TPCs for WIMP detection, with a
micro-TPC where the readout is performed by a microdots structure.

The field of WIMP direct detection is going through a phase in
which a great diversity of techniques are being developed and
tested. This diversity is important, as it must clarify which
techniques will prevail in the next generation of experiments.
Currently the best exclusions are obtained by cryogenic hybrid
detectors, whose success is based on a powerful separation between
electron and nuclear recoils, keeping on a second plane the more
conventional strategies based on radiopurity and shielding. But if
the WIMP lies relatively far from the present sensitivity levels
(say, more than 3 orders of magnitude, for example) an important
scaling up from the present prototypes is required, which seems a
bigger challenge for cryogenic detectors than, for instance, Xenon
ones. At those larger exposures, neutron background will limit the
sensitivity and again shielding techniques will become important
in any case. The strategy of self-shielding, at reach for large,
position sensitive detectors may well be the successful next step.
In addition, in case a WIMP appears, as an irreducible background,
in these kind of detectors, strategies to positively identify it
must be prepared. Any of the three signatures being pursued will
be useful, and specially the directionality one.

\section{Axions searches}

Axion phenomenology\cite{Raffelt:1990yz,Turner:1989vc} depends
mainly on the scale of the PQ symmetry breaking, $f_a$. In fact,
the axion mass is inversely proportional to $f_a$, as well as all
axion couplings. The proportionality constants depend on
particular details of the axion model considered and in general
they can be even zero. An interesting exception is the coupling
axion-photon, which arises in every axion model from the necessary
Peccei-Quinn axion-gluon term. This interaction is
phenomenologically described by:

\begin{equation}\label{lag}
    L_{a\gamma\gamma} = -g_{a\gamma\gamma} \phi_a \mathbf{E}\cdot\mathbf{B}
\end{equation}

\noindent where $\phi_a$ is the axions field, $\mathbf{E}$ and
$\mathbf{B}$ the electric and magnetic fields, and
$g_{a\gamma\gamma}=\alpha C_{\gamma}/\pi f_a$ the axion-photon
coupling, being $\alpha$ the fine structure constant and
$C_\gamma$ is a model dependent constant, usually of order unity
(for example, $C_{\gamma}\sim0.97$ for the
KSVZ\cite{Kim:1979if,Shifman:1979if} model and
$C_{\gamma}\sim-0.36$ for the
DFSZ\cite{Dine:1981rt,Zhitnitsky:1980tq} model).

The only axion phenomenology assumed in all experiments presented
in the next paragraphs is this axion-photon coupling. This
coupling allows for the conversion of axion into photons in the
presence of (electro)magnetic fields, a process usually called
Primakoff effect and that is beneath all the detection techniques
described in the following.

\subsection{Galactic axions}

Axions could be produced at early stages of the Universe by the
so-called misalignment (or realignment)
effect\cite{Turner:1989vc}. Extra contributions to the relic
density of non-relativistic axions might come from the decay of
primordial topological defects (like axion strings or walls).
There is not a consensus on how much these contributions account
for, so the axion mass window which may give the right amount of
primordial axion density (to solve the dark matter problem) spans
from $10^{-6}$ eV to $10^{-3}$ eV. For higher masses, the axion
production via these channels is normally too low to account for
the missing mass, although its production via standard thermal
process increases. Thermal production yields relativistic axions
(hot dark matter) and is therefore less interesting from the point
of view of solving the dark matter problem, but in principle axion
masses up to $\sim 1$ eV, are not in conflict with cosmological
observations\cite{Hannestad:2005df}.

The best technique to search for low mass axions composing the
galactic dark matter is the microwave cavity originally proposed
in\cite{Sikivie:1983ip}. In a static background magnetic field,
axions will decay into single photons via the Primakoff effect.
The energy of the photons is equal to the rest mass of the axion
with a small contribution from its kinetic energy, hence their
frequency is given by $hf = m_ac^2(1 + O(10^{-6}))$. At the lower
end of the axion mass window of interest, the frequency of the
photons lies in the microwave regime. A high-Q resonant cavity,
tuned to the axion mass serves as high sensitivity detector for
the converted photons.

The Axion Dark Matter Experiment
(ADMX)\cite{Asztalos:2001tf,Asztalos:2003px} has implemented the
concept using a cylindrical cavity of 50 cm in diameter and 1 m
long. The $Q$ is approximately $2\times10^5$ and the resonant
frequency (460 MHz when empty) can be changed by moving a
combination of metal and dielectric rods. The cavity is permeated
by a 8 T magnetic field to trigger the axion-photon conversion,
produced by a superconducting NbTi solenoid.

\begin{figure}[t]
\centering\mbox {\epsfig{file=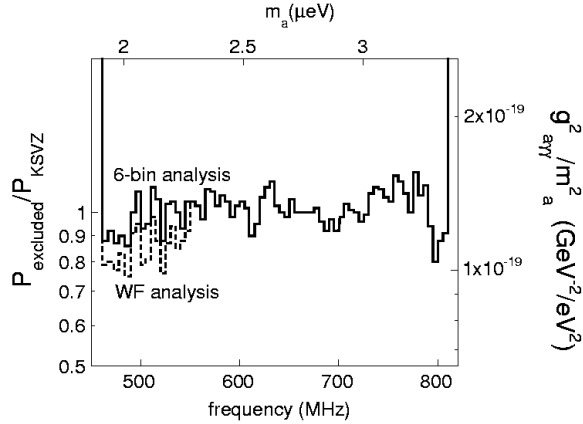,width=8.cm}} \caption{Upper
limit on axion-to-photon conversion power and coupling
$g_{a\gamma\gamma}$, excluded at greater than 90\% confidence,
assuming axion halo density 0.45 GeV/cm$^3$.} \label{admx}
\end{figure}

So far the ADMX experiment has scanned a small axion mass energy,
from 1.9 to 3.3 $\mu$eV\cite{Asztalos:2003px} with a sensitivity
enough to exclude a KSVZ axion, assuming that thermalized axions
compose a major fraction of our galactic halo ($\rho_a=450$
MeV/c$^2$). An independent, high-resolution search channel
operates in parallel to explore the possibility of fine-structure
in the axion signal\cite{Duffy:2005ab}. The detailed exclusion is
shown in fig. \ref{admx}.

Current work focuses on the upgrade of the experimental set-up,
which means basically to reduce the noise temperature of the
amplification stage. This is being done by newly developed SQUID
amplifiers and in a later stage by reducing the temperature of the
cavity from the present 1.5 K down to below 100 mK by using a
dilution refrigerator. These improvements will allow ADMX to
increase the sensitivity to lower axion-photon coupling constants
and also to larger axion masses.

\subsection{Solar axions}

Axions or other hypothetical axion-like particles with a
two-photon interaction can also be produced in the interiors of
stars by Primakoff conversion of the plasma photons. This axion
emission would open new channels of stellar energy drain.
Therefore, energy loss arguments constrain considerable axion
properties in order not to be in conflict with our knowledge of
solar physics or stellar evolution\cite{Raffelt:1999tx}.

In particular, the Sun would offer the strongest source of axions
being a unique opportunity to actually detect these particles. The
solar axion flux can be estimated \cite{vanBibber:1989ge,Creswick}
within the standard solar model. The expected number of solar
axions at the Earth surface is $\Phi_a=(g_{a\gamma}/10^{-10}\, \rm
GeV^{-1})^2\,3.54\times10^{11}~\rm cm^{-2}~s^{-1}$ (being
$g_{a\gamma}$ the axion-photon coupling) and their energies follow
a broad spectral distribution around $\sim$4~keV, determined by
solar physics (Sun's core temperature). Solar axions, unlike
galactic ones, are therefore relativistic particles.

%

These particles can be converted back into photons in a laboratory
electromagnetic field. Crystalline detectors may provide such
fields \cite{Paschos:1993yf,Creswick:1997pg}, giving rise to very
characteristic Bragg patterns that have been looked for as
byproducts of dark matter underground experiments
\cite{Avignone:1997th,Morales:2001we,Bernabei:ny}. However, the
prospects of this technique have been proved to be rather limited
\cite{Cebrian:1998mu}, an do not compete with the experiments
called ''axion helioscopes'' \cite{Sikivie:ip,vanBibber:1989ge},
which use magnets to trigger the axion conversion. This technique
was first experimentally applied in \cite{Lazarus:1992ry} and
later on by the Tokyo helioscope~\cite{Moriyama:1998kd}, which
provided the first limit to solar axions which is
"self-consistent", i.e, compatible with solar physics. Currently,
the same basic concept is being used by the CAST collaboration at
CERN \cite{Zioutas:1998cc,Andriamonje:2004hi} with some original
additions that provide a considerable step forward in sensitivity
to solar axions.

The CAST experiment is making use of a decommissioned LHC test
magnet that provides a magnetic field of 9 Tesla along its two
parallel pipes of 2$\times$14.5 cm$^2$ area and 10 m long. These
numbers mean that the axion-photon conversion probability is a
factor 100 higher than in the previous best helioscope at Tokyo.
The CAST magnet has been mounted on a platform that allows to
point it to the Sun and track it during $\sim$3 h per day in
average. The rest of the day is devoted to measure the background
experimentally. CAST adds up expertise in low background
techniques to operate three different X-ray detectors with
complementary approaches: a TPC, a MICROMEGAS and a CCD. A
relevant component of the experiment is the X-ray focussing mirror
system, designed and built as a spare system for the X-ray
astronomy mission ABRIXAS, and now recovered for CAST. It provides
a focussing of the X-rays coming out of the magnet down to a spot
of a few mm$^2$ on the CCD, further increasing the signal-to-noise
ratio and therefore the sensitivity of the experiment.

\begin{figure}[t]
\begin{center}
\includegraphics[width=85mm]{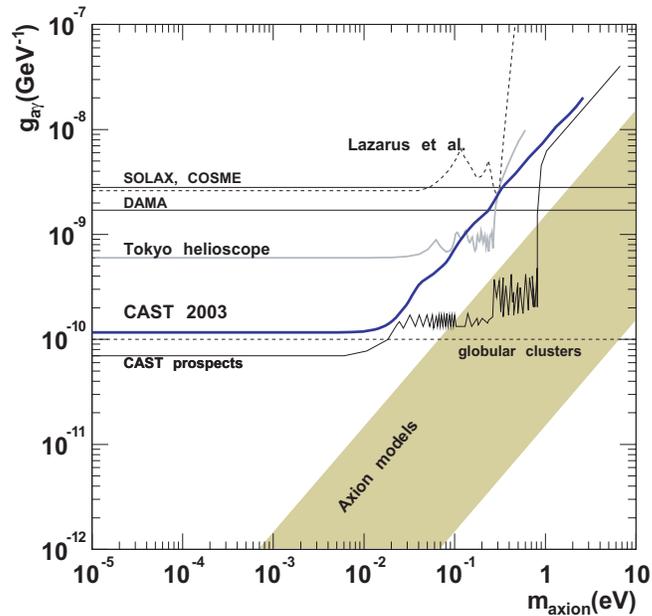}
 \caption{95\% CL exclusion line obtained from the analysis
 referred to
 in the present paper (line labeled "CAST 2003"), compared with other
 limits.
 \label{exclusion}}
 \end{center}
\end{figure}

CAST has been running in 2003 and, in improved conditions, in
2004. The results of the analysis of the 2003 data have been
recently released\cite{Andriamonje:2004hi}. No signal above
background was observed, implying an upper limit to the
axion-photon coupling $g_{a\gamma} < 1.16 \times 10^{-10}~{\rm
GeV}^{-1}$ at 95\% CL for the low mass (coherence) region
$m_a\lsim 0.02~{\rm eV}$. As can be seen in figure \ref{exclusion}
this limit is a factor 5 more restrictive than the limit from the
Tokyo axion helioscope and already comparable to the one derived
from stellar energy-loss arguments. The 2004 data will allow to
improve the sensitivity of the experiment close to the expectation
of the experiment proposal. Currently the experiment is being
adapted for the second phase which consist in data taking with a
buffer gas (He$^4$ and/or He$^3$) inside the magnet pipes. Varying
the pressure of the gas allows to match the coherence condition
for a range of higher axion masses up to $\sim$eV. As can be seen
in figure \ref{exclusion}, CAST phase II sensitivity will enter
for the first time the region of the axion parameter space where
the most theoretically motivated axion models lie.

\subsection{Laboratory axions}

The existence of axions or other axion-like particles may produce
measurable effects in the laboratory. A typical example is the
"light through wall" experiments, in which a photon beam is
converted into axions inside a magnetic field and, after crossing
an optical barrier, are converted back into photons by another
magnetic field. As a result, light seems to have gone through an
opaque wall. This technique was used to derive some early limits
on the axion properties \cite{Eidelman:2004wy}.

A more subtle effect is the magnetic-induced birefringence of the
vacuum. When a polarized photon beam traverses an empty space
permeated with a magnetic field, the polarization component
parallel to the magnetic field gets out-of-phase with respect the
perpendicular one, producing an \emph{ellipticity} on the final
polarization of the beam. Such an effect is predicted by standard
QED by virtue of four-legged fermion loops, although its magnitude
is extremely small. Experiments with ultra-precise optical
equipment may look for such an effect. The PVLAS
experiment\cite{pvlas}, designed to measure the QED-predicted
magnetic-induced birefringence has been systematically detecting
such a signal, however 4 orders of magnitude larger than expected.
After several years of tests to rule-out possible systematic
effects, the collaboration has officially announced that they are
indeed detecting an unusually large unexplained ellipticity in
their laser beam\cite{pvlas} connected with the presence of the
magnetic field. One tentative explanation might be the presence of
an photon-axion oscillation. Such scenario predicts a second
effect which would distinguish it from the standard QED effect. It
consists of a rotation of the polarization (or \emph{dichroism})
due to the reduction of one the polarization components, produced
by real photon-axion conversion. The presence of such additional
effect has been recently confirmed also by PVLAS. However, the
interpretation of PVLAS observation in terms of axions needs an
axion mass of $\sim 1$ meV and an axion-photon coupling of $\sim
10^{-6}$ GeV$^{-1}$, far larger than present experimental and
solar limits\cite{pvlas} (although exotic extensions of the
standard axion scenario may allow to reconcile all experimental
results\cite{Masso:2005ym}). In any case, the nature of the PVLAS
effect is still an open question.

\section{Conclusions}

A review of the current status of the experimental searches for
WIMPs and axions has been given. The field lives a moment of great
activity, triggered by the fact that very well motivated
theoretical candidates could be within reach of present
technologies. The presence of intriguing positive signals both in
WIMPS and axion detection has at least enhanced the interest and
excitation of the research. The next years will witness the
results of many very interesting developments currently ongoing to
define and operate a new generation of experiments, well into the
region of interest for both WIMPs and axions.

\section{Acknowledgements}

I would like to dedicate the present review to the memory of prof.
A. Morales. I am indebted to him for many years of collaboration
and tutoring in the field of direct searches for dark matter. He
used to give very detailed and thorough reviews on dark matter in
the most important international conferences. His last review, in
the TAUP2003 at Seattle\cite{Morales:2005ky} has been particularly
useful in the WIMP section of the present review.


\end{document}